# Downlink and Uplink Decoupling: a Disruptive Architectural Design for 5G Networks

Hisham Elshaer[*#], Federico Boccardi[*], Mischa Dohler[#] and Ralf Irmer[*]
[*]Vodafone Group R&D. Newbury, UK
[#]King's College London. London, UK
Email: (hisham.elshaer, federico.boccardi, ralf.irmer)@vodafone.com
mischa.dohler@kcl.ac.uk

*Abstract*—Cell association in cellular networks has traditionally been based on the downlink received signal power only, despite the fact that up and downlink transmission powers and interference levels differed significantly. This approach was adequate in homogeneous networks with macro base stations all having similar transmission power levels. However, with the growth of heterogeneous networks where there is a big disparity in the transmit power of the different base station types, this approach is highly inefficient. In this paper, we study the notion of Downlink and Uplink Decoupling (DUDe) where the downlink cell association is based on the downlink received power while the uplink is based on the pathloss. We present the motivation and assess the gains of this 5G design approach with simulations that are based on Vodafone's LTE field trial network in a dense urban area, employing a high resolution ray-tracing pathloss prediction and realistic traffic maps based on live network measurements.

## I. INTRODUCTION

In order to keep up with the ever increasing network traffic, cellular networks are shifting from a single-tier homogeneous network approach to multi-tier heterogeneous networks (HetNets). HetNets, composed of different types of small cells (Micro, Pico and Femto), have been a popular approach in the past few years as an efficient and scalable means to improve the network capacity in hotspots. However, most network technologies such as 3G or 4G were designed with Macro cells in mind and heterogeneity was just an afterthought. This dramatic change in cellular networks requires a fresh look on how present networks are deployed and what fundamental changes and improvements need to be done for future networks to operate efficiently.

Cellular networks have often been designed based on the downlink (DL); this is due to the fact that network traffic is mostly asymmetric in a way that the throughput required in the downlink is higher than the one required in the uplink. However, uplink is becoming more and more important with the growth of sensor networks and machine type communications (MTC) where the traffic is often uplink centric and also the increasing popularity of symmetric traffic applications, such as social networking, video calls, real-time video gaming, etc. As a consequence, the optimization of the uplink has become increasingly important and the question that we try to tackle in this paper is what improvements are needed to optimize the uplink of a highly densified HetNet?

Cell association in cellular networks is normally based on the downlink received signal power only [1]. Despite differing UL and DL transmission powers and interference levels, this approach was sufficient in a homogeneous network where all the BSs are transmitting with the same or similar average power level. However, in HetNets where we have a large disparity in the transmit power of the different layers this approach is highly inefficient in terms of the uplink.

To understand this assertion we consider a typical HetNet scenario with a Macro cell (Mcell) and a small cell (Scell), where in this paper we consider outdoor small cells. The DL coverage of the Mcell is much larger than the Scell due to the large difference in the transmit powers of both. However, in the UL all the transmitters, which are battery powered mobile devices, have about the same transmit power and thus the same range. Therefore, a user equipment (UE) that is connected to a Mcell in the DL from which it receives the highest signal level might want to connect to a Scell in the UL where the pathloss is lower to that cell.

As HetNets become denser and small cells smaller, the transmit power disparity between macro and small cells is increasing and, as a consequence, the gap between the optimal DL and UL cell boundaries increases. For the sake of optimal network operation, this necessitates a new design approach which is the Downlink and Uplink Decoupling (DUDe) where the UL and DL are basically treated as separate network entities and a UE can connect to different serving nodes in the UL and DL.

### A. Related Work

The concept of DUDe has been discussed as a major component in future cellular networks in [2]-[4]. In [4], in particular, DUDe is considered as a part of a broader "device-centric" architectural vision, where the set of network nodes providing connectivity to a given device and the functions of these nodes in a particular communication session are tailored to that specific device and session. A study in [5] tackles the problem from an energy efficiency perspective where the

UL/DL decoupling allows for more flexibility in switching-off some BSs and also for saving energy at the terminal side. In [6] Multi-Radio HetNets are discussed where all radio access technologies (RAT) like WiFi and LTE are managed under a single network and this can be considered as an extension to DUDe in future work where UL and DL can be scheduled on different RATs.

One technique that brings some fairness to the UL is "Range Extension" (RE) where the idea is to add a cell selection offset to the reference signals of the Scells to increase their coverage in order to offload some traffic from the Mcells [7]. However, using offsets greater than 3-6 dB may lead to high interference levels in the DL which is why techniques – like enhanced Inter-Cell Interference Coordination (eICIC) – have been developed to try to combat this type of interference [8]. Nevertheless, the RE technique is limited to moderate offset values due to the harsh interference in the DL. So DUDe would bring in the benefits of having very high RE offsets in the UL without the interference effects in the DL.

B. Contribution

The main contribution of this paper is to study the gains that can be achieved by the DUDe technique in terms of UL capacity and throughput and also to study the effects that this approach has on interference. We use a realistic scenario of a cellular network based on real-world planning/optimisation tools which, we believe, adds a lot of value and credibility to this work. In our best knowledge, this is the first work that assesses the benefits of decoupling UL and DL in a real world deployment.

The rest of the paper is organized as follows: in Section II we present the gains and motives of DUDe based on a simplified network model. In Section III we present the simulation setup. Section IV includes the simulation results and finally in Section V we present the conclusions and final remarks.

## II. SIMPLIFIED MODEL

In this study we drop the traditional UL/DL cell association based on DL received power (RP). We assume that while the downlink association is still based on DL RP, the uplink association is in fact based on pathloss. This apparently simple assumption in reality leads to radical changes in system design and architecture. One issue with this approach is when a UE has a link in 1 direction to a node (UL or DL) it needs a mechanism to allow the Acknowledgment process, channel estimation, etc. This would require major design changes. Therein we aim at studying whether the gains of DUDe justify such major changes.

DUDe results in different cell boundaries in the UL and DL in a HetNet scenario where a UE in the region between the UL and DL cell boundaries will be connected to the Scell and Mcell in the UL and DL respectively as shown in Figure 1. We will focus on the gains in the UL as this is the main motive for applying this technique. Note that DL capacities are not affected since the association remains unchanged.

In this section, we consider a two cell network model composed of a Mcell and a Scell to present the advantages of DUDe in a simplified way. The model is used to study two cases; the first case is a noise limited scenario with only one UE, to show the benefits in terms of uplink UE capacity. The second case is an interference limited scenario where there are three UEs in the network to show the benefits in reducing the interference. The two cases are explained in details below.

A. Case 1 (noise limited)

In this case we have 1 UE moving from the Scell vicinity towards the Mcell and the UE UL rate is calculated for two cases; the first is the conventional case where cell selection is based on the DL received power so the UE performs a Handover (UL & DL) from the Scell to the Mcell when passing the DL cell border (shown in Figure 1) and the second case is where the UL cell selection is based on the PL where the UE is still connected to the Scell until passing the UL cell border which represents the DUDe technique.

Neglecting, for simplicity, fading and shadowing and normalizing various quantities, the UL rate calculation is based on the below equations:

$$R = BW \log_2(1 + SNR)$$

$$SNR = \frac{P_{ue}}{N\,d^\alpha}$$

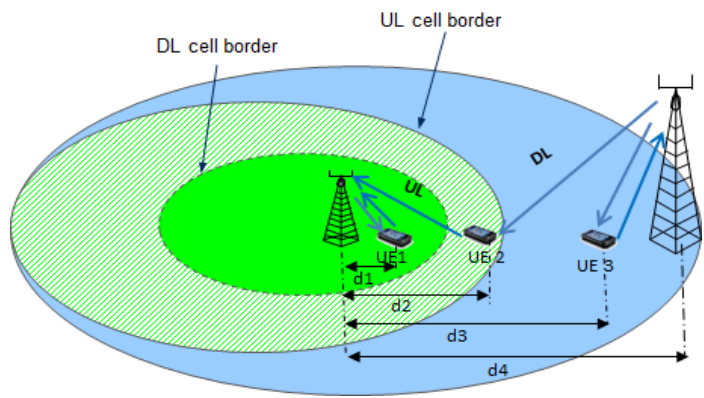

Figure 1. System model for the UL/DL decoupling.

Here, $R$ is the rate; $SNR$ is the signal to noise ratio, $P_{ue}$ is the UE transmit power and N is the noise power which is considered to be 0 dBm. $BW$ is the bandwidth and is considered to be unity for simplicity. The distance based $PL$ is dependent on the distance $d$ and the pathloss exponent $\alpha$.

We now calculate the UL rate for a UE moving from the Scell towards the Mcell for the two cell association methods, assuming, $P_{ue}$ to be 20 dBm and the Scell and Mcell to have a PL exponent of 3.6 and 4 respectively. Finally, the Mcell and Scell have a transmit power of 46 and 23 dBm respectively.

Figure 2 illustrates the UL normalized rate for the PL and RP cell association cases. And it shows that the PL case has a higher performance in the area between the DL cell border and the UL cell border since in that area the UE has a lower pathloss to the Scell, thus obtaining a higher rate when connected to the Scell. The two curves are the same outside that area since the PL and RP cell association result in selecting the same cell.

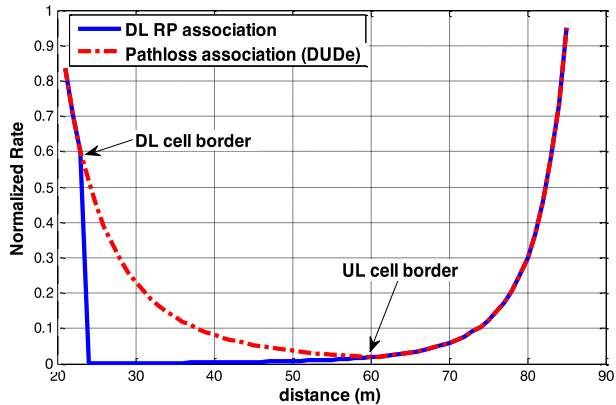

Figure 2. UE Rate comparison between the DL RP case and the PL case.

B. Case 2 (interference limited)

In this case, we have the same setup as the previous one but with three UEs instead of only one UE as shown in Figure 1. We calculate the overall UL Rate of the network using the PL based cell association where UE2 is connected to the Scell in the UL and then using the RP based cell association where UE2 is connected to the Mcell in the UL.

UE1 is always connected to the Scell in the UL and UE3 is always connected to the Mcell in the UL.

$$R = BW \, \log_2(1 + SIR)$$

$$SIR = \frac{P_{ue}}{I \, d^\alpha}$$

The UL Rate is calculated based on the above equation, where SIR is the Signal to Interference Ratio (we neglect the noise for simplicity). The total normalized UL rate ($R_T$) is the sum of the normalized UL rate at the Mcell ($R_M$) and the Scell ($R_S$) which means the UL rate of the whole system ($R_T = R_M + R_S$).

We use the same parameters as case 1 and setting d1, d2, d3, and d4 in Figure 1 to 10, 25, 80, and 100 respectively. So calculating $R_T$ in the PL case yields $R_T = 0.46 + 0.54 = 1$ and in the RP case $R_T = 0.34 + 0.33 = 0.67$. We can see that $R_T$ is 50% higher in the PL case for the following reasons.
- UE2 in the PL case has a lower PL to the Scell which means that UE2 has a better channel to the Scell and in turn gets a better Rate when connected to it.
- UE2 causes less interference to the Mcell in the PL case than the interference it causes to the Scell in the RP case for the same reason as above, so the interference level in the network is lower and in turn the rate is higher.

In the next section we present our realistic simulation setup which is based on an existing cellular network and we use this setup to validate our findings and illustrate the gains from the studied concept.

III. SIMULATION SETUP

In our simulations we use the Multi-technology radio planning tool Atoll [9] in conjunction with a high resolution 3D ray tracing pathloss prediction model [10]. The model takes into account clutter, terrain and building data. This guarantees a realistic and accurate propagation model.

Atoll has the capability of performing system level simulations where a simulation is a snapshot of the LTE network. For each simulation, it generates a user distribution using a Monte Carlo algorithm. The user distribution is based on traffic data extracted from the real network and is weighted by a Poisson distribution. Resource allocation in each simulation is carried out over a duration of 1 second (100 frames).

As deployment setup, we use a Vodafone LTE small cell test bed network that is up and running in the London area. The test network covers an area of approximately one square kilometer. We use this existing test bed to simulate a relatively dense HetNet scenario.

The considered network is shown in Figure 3 where the black shapes are Macro sites and the red circles are Small cells which are considered to be Pico cells.

We consider a realistic user distribution based on traffic data from the field trial network in peak times. The distribution is up-scaled to simulate a high user density.

We use an uplink power control algorithm where each cell has a predefined interference upper limit. If the UL received interference at a cell is higher than this limit the cell signals the neighboring cells to lower the UL transmit power of their UEs to lower the interference level at that cell.

We simulate two cases; the first case is where the UL cell association is based on PL which represents the DUDe technique. The other case is where the UL cell association is based on the DL Reference Signal Received Power (RSRP) which is the conventional LTE procedure [1]. In the DL RSRP case we simulate low and high power Scell cases to understand the gains of the PL approach compared to the DL RSRP approach with different Scell sizes.

As pointed out before, all the results in the next section will focus on the UL performance. The simulation parameters are listed in Table 1 where we consider an LTE deployment.

One deployment issue is that a UE connected to different nodes in the UL and DL needs a way to send Acknowledgment, pilot and relevant control signaling to its DL node with which it has no UL established. A possible way is to route the data to the UL node and through the backhaul to the DL node and vice versa with receiving control signals from the UL node.

We assume an ideal backhaul where control signals are delivered with no notable delay. Non-ideal backhaul operation and alternative control signaling delivery mechanisms are left for future work.

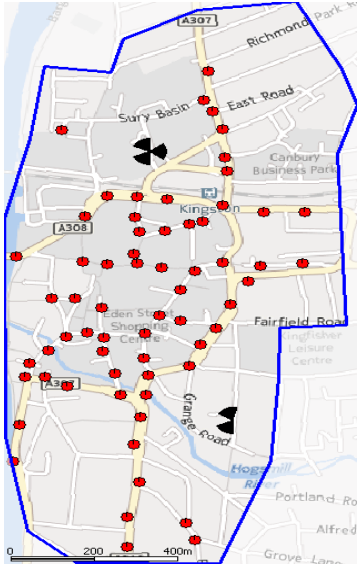

Figure 3. Vodafone small cell LTE test network in London.

Table 1. Simulation Parameters

| | |
|---|---|
| Operating frequency | 2.6 GHz (co-channel deployment) |
| Bandwidth | 20 MHz (100 frequency blocks) |
| Network deployment | 5 Mcells and 64 Scells distributed in the test area as shown in Figure 3. |
| User distribution | 560 UEs distributed according to traffic maps read from a live network |
| Scheduler | Proportional fair |
| Simulation time | 50 simulation runs with 1 second each. |
| Traffic model | Full buffer |
| Propagation model | 3D ray-tracing model |
| Max. transmit power | Macro=46 dBm, High power Pico = 30dBm, Low power Pico = 20 dBm, UE= 20 dBm. |
| Antenna system | Macro: 2Tx, 2Rx, 17.8 dBi gain Pico: 2Tx, 2Rx, 4 dBi gain UE: 1Tx, 1Rx, 0 dBi gain |
| UEs mobility | Pedestrian (3km/h) |
| Supported UL modulation schemes | QPSK, 16 QAM, 64 QAM |

## IV. SIMULATION RESULTS

In this section, we present results comparing three cases:
- DL RP based cell association where Scells are Pico cells (Pcells) with low transmit power (LP) of 20 dBm. This case is referred to as DL-LP.
- DL RP based cell association where Scells are Pcells with high transmit power (HP) of 30 dBm. This case is referred to as DL-HP.
- Pathloss based cell association which represents the Downlink and Uplink Decoupling (DUDe) (Pico transmit power is irrelevant as cell association is not based on DL RP).

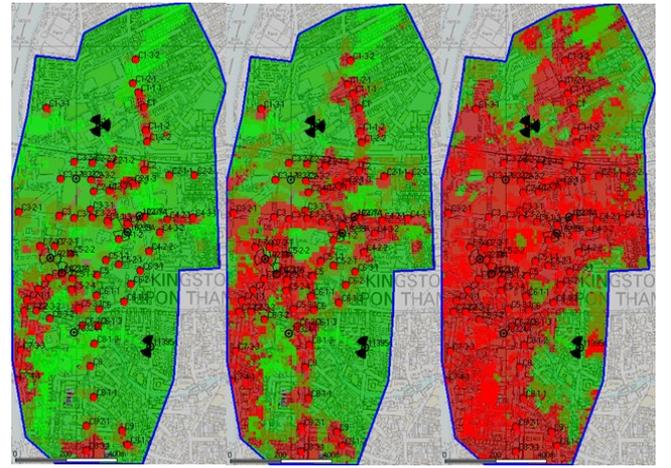

Figure 4. Uplink coverage of the DL_LP (left), DL_HP (middle) and DUDe (right) cases where green and red represent the Macro and Pico cells coverage respectively.

Figure 4 illustrates the UL coverage of the Pcell layer (red) and Mcell layer (green) for the above three cases; it shows a much larger coverage for the Pcells in the DUDe case which ensures a more homogeneous distribution of UEs between the nodes which, in turn, results in a much more efficient use of the resources as will be shown in the following results.

In our simulations we define a UE minimum and maximum throughput demand where basically a UE has to reach the minimum throughput requirement to be able to transmit its data otherwise it is considered in outage. On the other hand the maximum throughput demand puts a limit to the amount of throughput that each UE can get, so setting a high value for it helps in simulating a highly loaded network.

The used scheduler tries first to satisfy the minimum throughput requirements for all the UEs and then distributes the remaining resources among the UEs to satisfy the maximum throughput demand of each UE according to the proportional fair criterion.

Figure 5 shows the effect of adding Pico cells on the $5^{th}$ percentile UL throughput for the different cases. Pcells are all placed in their respective location as shown in Figure 3 but they are all switched off at the beginning and are activated one by one to understand the effect of increasing the number of Pcells in each case. In these results, we set the minimum throughput requirement to a relatively low value (100Kb/s) to show how the $5^{th}$ percentile throughput evolves in the different cases without the constraint of a high minimum throughput requirement.

In the DUDe case, we see that the $5^{th}$ percentile throughput is increasing with the number of Pcells. This is due to the fact that Pcells have a large coverage in the UL so they serve a large number of UEs and in turn have a big effect on the $5^{th}$ percentile throughput. As the number of Pcells increases we notice that the $5^{th}$ percentile UEs throughput starts to saturate as they are more limited by the channel quality and transmit power. So the extra capacity offered by adding more Pcells is used to serve the UEs with better channel conditions. Looking at the case of DL-LP and DL-HP, we see that adding Pcells has little effect on the 5th percentile throughput as Pcells have very limited coverage so their effect is more in the 90th percentile throughput rather than the 5th percentile.

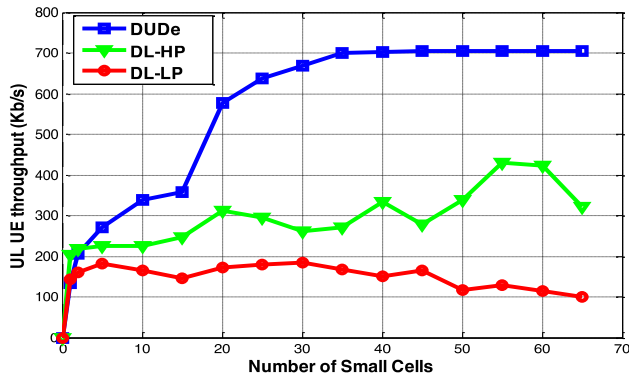

Figure 5. The 5th percentile UL UE throughput, comparing the DUDe, DL-HP and DL-LP cases with increasing the number of Pico cells.

Moreover; we see that the 5th percentile throughput is fluctuating as we increase the number of Pcells. This is basically due to the high interference that the Pcells UEs create to the Mcell cell edge UEs since these UEs are closer to the Pcells so they suffer from a high level of interference. We see this effect more clearly in the DL-LP case where the throughput starts to decrease after a certain point. On the contrary, in the DUDe case we see that the throughput is increasing more stably since the UEs always connect to the node to which they have the lowest PL which guarantees a lower interference level as explained in Section II. In the next results all the Pcells in the test network are activated.

Table 2. Average number of UEs per Node (Macro and Pico cells) for the three cases.

|  | DL-LP | DL-HP | DUDe |
|---|---|---|---|
| Macro cell | 81 | 50 | 13 |
| Pico cell | 2 | 4 | 8 |

Table 2 shows the average number of UEs per cell where we calculate the average for the Mcells and Pcells separately having a constant total number of UEs (560) for all cases. The table shows how most of the UEs are connected to the Mcell in the DL-LP and DL-HP cases and the Pcells are under-utilized. On the other hand in the DUDe case the UEs are distributed in a more homogeneous way among the Mcells and Pcells which ensures much more efficient resource utilization.

For the results in Figure 6 we set a minimum and maximum throughput demand of 200 Kb/s and 20 Mb/s respectively. The figure shows the 5th, 50th, and 90th percentile UE throughput for the three cases in comparison. The 5th percentile UL throughput in the DUDe case is increased by more than 200% compared to the DL-LP case and by 100% compared to the DL-HP.

As for the 50th percentile UL throughput, the DUDe case has a gain of more than 600% compared to the DL-LP case and more than a 100% compared to the DL-HP case.

The gains in the 5th and 50th percentile are resulting from the higher coverage of the Pcells in the DUDe case which results in a better distribution of the UEs among the nodes and a much more efficient usage of the resources.

Also the fact that the UEs connect to the node to which they have the lowest PL helps in reducing the UL interference as shown before. This results in a higher UE Signal to Noise and Interference Ratio (SINR) that allows the UEs to use a higher modulation scheme and in turn achieve a better utilization of the resources and a higher throughput.

Looking at the 90th percentile UL throughput we see that the DL-HP case achieves the highest throughput which can be explained by the fact that Pcells serve less UEs than the DUDe case then these UEs get a high throughput but on the expense of the 5th and 50th percentile UEs. Interestingly the DL-HP case achieves a higher 90th percentile throughput than the DL-LP case which seems counter intuitive. This can be explained by the fact that Pcells in the DL-LP case serve even less UEs than the Pcells in the DL-HP case so the effect of the Pcells in the DL-LP case is noticeable even after the 90th percentile. So if we look at the 98th percentile throughput in the DL-LP case it's 15 Mb/s whereas in the DL-HP case it's 10 Mb/s which shows that the effect of the Pcells in the DL-LP case is on a very limited number of UEs.

The gains of the 5th and 50th percentiles are comparable to the results shown in [11] where the authors apply a high Range Extension (RE) value to the Pcells and they get a two times gain in the 5th and 50th percentile throughput. The RE technique basically works in the same direction as decoupling the UL and DL in the sense that it results in an increased coverage in the UL. The disadvantage of RE is that the interference level in the DL increases aggressively as the RE bias increases which requires the usage of interference management techniques as mentioned before which is not required in the UL/DL decoupling since the UL and DL are treated as two different networks in this technique.

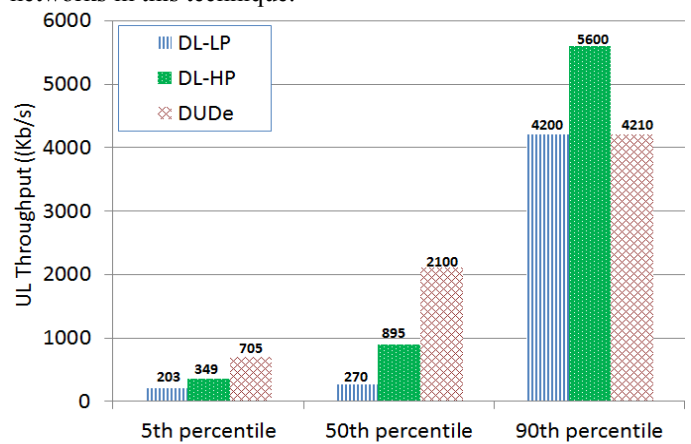

Figure 6. 5th, 50th and 90th percentile comparison of DL-LP, DL-HP and DUDe cases.

In the next result we run the same simulation but increasing the minimum throughput demand to 1 Mb/s to study the outage rate in this high density scenario.

Figure 7 represents the average outage rate for the Mcell layer and the Pcell layer for the three cases or in other words the percentage of the UEs that fail to achieve the minimum throughput demand (1 Mb/s) out of the total number of connected UEs to a certain node. Since the simulated scenario is considered to be a highly dense scenario it requires a very efficient use of resources in order to satisfy the high requirements of the UEs. As seen in the figure the Macro layer has a very high outage rate (more than 90%) in the DL-HP and DL-LP cases which is basically explained by the fact that the Macro layer is very congested in the UL as seen in Table 2 so Mcells do not have enough resources to serve all their UEs with a high throughput level. However, in the DUDe case, UEs are distributed more evenly between the nodes so the outage rate in that case is low (less than 10%) for both Macro and Pico cell layers.

These results clearly show that decoupling UL and DL where UL is based on PL is a promising candidate for future networks where the network load is expected to increase in the UL and where providing a consistent and ubiquitous service to all UEs in different network deployments and UE densities is a priority. This technique would also allow freeing up spectrum resources in the UL which could be used for DL purposes.

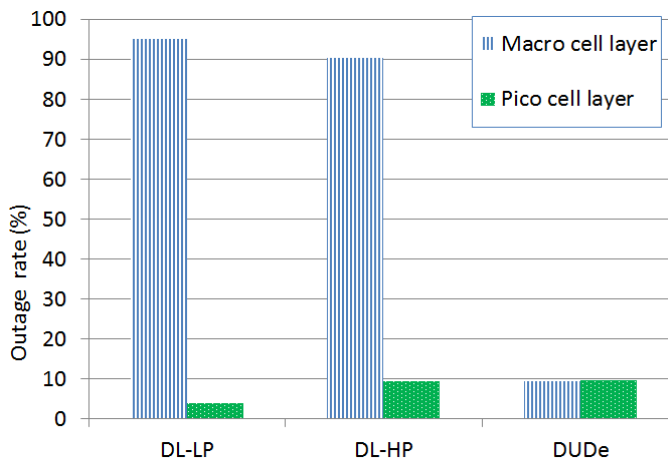

Figure 7. Average outage rate of Macro and Small cell layers comparison between the DL-LP, DL-HP and DUDe cases.

## V. CONCLUSIONS

In this paper, we presented an assessment of the UL/DL decoupling concept (DUDe) in a dense HetNet deployment. We started by a simplified model to highlight the motives and gains of this concept and then we presented simulation results based on a live Vodafone LTE test network deployment in London. The simulations used a high resolution ray tracing propagation model and user distributions based on network measurements which make this model highly realistic and providing a much better view on the effects of deploying this technique in the real world than normal system level simulations. The gains are quite high in a dense HetNet deployment where this technique can achieve two to three times better $5^{th}$ percentile UL throughput and even more than that in the $50^{th}$ percentile throughput. Also, we have shown that the outage rate is decreased dramatically in networks with high minimum throughput requirements. We believe that the DUDe technique is a strong candidate for 5G architecture designs and it can be very useful in many applications like Machine Type Communications (MTC) where Uplink optimization is very critical. Our future work will focus on studying the network architectural and design changes that would enable the decoupling of UL and DL.

ACKNOWLEDGMENT

This research has been co-funded by Vodafone Group R&D and CROSSFIRE (un-CooRdinated netwOrk StrategieS for enhanced interFerence, mobIlity, radio Resource, and Energy saving management in LTE-Advanced networks) MITN Marie Curie project. The authors would like to thank E. Murray, N. Scully, J. Turk, I. Thibault and E. Bouton for their input and useful feedback.